# Traffic control Management System and Collision Avoidance System

**Gangadhar** [1] **Dr.Parimala Prabhakar** [2]**,** ***Abhishek S*** [3] ***, Prajwal***[4]***, Suraj Naik*** [5]

[1345]*Dept. of Electronics and Telecommunication Engineering, Ramaiah Institute of Technology,* Bangalore, India
[2]*Assistant Professor, Dept. of Electronics and Telecommunication Engineering, Ramaiah Institute of Technology, Bangalore, India*

gangadhar.ajapur2017@gmail.com[1] , parimalap@msrit.edu[2], abhisheksubramaniyam@gmail.com[3], prajwalmallapur123@gmail.com [4], surajnaik811@gmail.com[5]

**Abstract: Many road accidents occur due to drivers failing to read sign board due to various reasons. Especially at night, the tiredness of driver reduces his perception to small things like speed limit of sign the board, curve ahead sign board. For the smooth movement of ambulance in cities during traffic, is to create an IOT device to detect sign boards and also able to communicate with the traffic light and makes way for ambulance. Implementation is done by detecting sign boards and measuring speed of vehicle using arduino and RF transmitter which transmits the specific beep sound to specific type of application like speed breaker, school zone etc. The vehicle also contains RF receiver and arduino, which starts receiving the beep sound when near to sign board. After receiving the code, arduino starts measuring the current speed of vehicle and if the speed is above recommended speed then it starts gives alert. If the vehicle speed is not reduced even after the alert then the vehicle will auto break. With the help of this Traffic Management System (TMS), we can record the number of users who do not reduce vehicle speed even when prompted by the system alerts.**

## 1 INTRODUCTION

In India the number of accidents per day is around 1,274 as per the report by NCBI, out of which many accidents occur due to driver not recognizing the sign boards.

At present the sign boards are used to alert the driver about the upcoming speed breaker, turnings, accident spots, school zone etc. But the high level of traffic and narrow roads in India makes driver distracted and driver fails to recognize the warning sign boards. By designing a system in which the vehicle should be able to recognize the sign boards even if the driver fails to recognize and alert the driver if proper action has not been taken within certain time interval, we can reduce the number of accidents. This feature can make them abide to rules and regulations.

In order to minimize and overcome the problems of road safety, we have developed an automatic Traffic Sign Detection and Recognition system or TSDR system. This system can be used for effectively detecting and recognizing different traffic signs from the captured images or by using image sensors. In certain situations, the driver may not notice the oncoming Traffic signs which may lead to fatal accidents. The TSDR system comes to use for such adverse situations. The main objective of the system is to develop the TSDR system to be an automatic system although it may be a difficult job owing to the fact that we have to take the continuous environmental and lighting conditions.

The other issues that play an important factor are the multiple traffic signs appearing at the same instant of time, blurred images of fading traffic signals. We must also make sure that the TSDR system should avoid the false detection of non-sign boards.

The aim of this project is to develop a system which can efficiently detect the traffic signs and classify the detected signs into separate classes. All of these actions are to be performed considering the real-time scenario.

## 2 RELATED WORK

Vehicle speed control and over speed violation using IOT [1], From this system we learnt that it can automatically control the speed of a particular vehicle by detecting speed sign labels from the nearby signboard. In case the driver does not slow down his vehicle speed, the vehicle details such as license plate number are sent to authorities and speed of the vehicle is reduced to the threshold level.

The smart accident detection and control system [2] From this system we could infer that it was performed on intersecting roads utilizing Raspberry Pi and microcontroller. The proposed system is capable of automatically detecting any vehicle accident. It then takes necessary steps to help the drivers who were injured due to the accident. The system also sends warning messages to nearby vehicles regarding the location of the road accident.

A smart system [3] has been proposed to alert and control speed of vehicle and also to notify individuals in the area in case of an accident. The system uses a distance sensor to monitor gap between vehicles or obstacles. Whenever an accident has occurred ,an email will be sent to the accountable driver with vehicle details.

This . GSM and GPS based system[4] was to effectively designed to prevent any kind of accidents and to detect them if such events occur. GSM and GPS were used to get the exact location of the user.

The signboard monitoring system [5] is used to detect when a vehicle meets with an accident. This is done with the help of vibration sensors that detect vibration signals in case of vehicle collisions and sends it to a Raspberry Pi controller. The microcontroller then sends mail alerts through IOT to the rescue team to help the driver who has met with an accident.

Detection of accidents precisely by means of both vibration sensor and Alcohol detection using eye blink sensors [6], As a future implementation to the existing system, we can add a wireless webcam which can be used for capturing the images of the accident site which will help in providing driver assistance.

Smart Accident Detection and Control System (SAD-CS) for intersecting road (‘Plus’ junction) [7], [8] This system can be developed by using Raspberry Pi Microcontroller. An on-board unit in every vehicle is required for such a system. The proposed system is capable of automatically and instantaneously detecting the vehicle accident and perform actions to help the affected people. The system also propagates warning messages to all the roadside vehicles to avoid the further accidents in the same lane. For accident avoidance implementation in the system, tire pressure of the vehicle is measured. The accident detection part of the system is implemented with the help of node MCU controller which has an inbuilt WiFi module. MQ7 sensor is used in order to monitor the pollution level in the particular traffic lane. This helps to reduce vehicular accidents as well as get information about the environmental status of the road by pollution monitoring.

The driver-assistance system [9] This system is capable of lane and painted traffic sign detection by using a car's on-board camera. This is integrated in a network which connects different users to enhance the efficiency of the detection of the painted traffic sign boards. A novel system to automatically control the speed of the vehicle [10] This system performs speed control of a vehicle by first detecting the speed signs labels from speed sign boards, which are laid on the roadside. Then the system takes necessary steps to take it to the knowledge of the driver by sending a caution notification indicating the driver to adjust the vehicle speed according to to the speed sign label that has been detected.

## 3 PROPOSED SYSTEM

The following are the main objectives of Traffic Management and Collision Avoidance using IoT:

•To design a smart sign board which can communicate with the vehicle.

•To send visual and audio alert to the vehicle if driver fails to adjust speed according to the detected sign board.

•To alert the user if the vehicle speed is not reduced if the detected sign board is a speed breaker/hump in the road.

•To develop an IR receiver-transmitter system for allowing ambulance to move in a particular road and making way for traffic.

•To record the number of vehicles that have been alerted by the buzzer to reach optimal speed in case of a sign board detected using IOT.

TMSCA has two functional units. One is for detection of sign board and other one is measuring speed of vehicle and user alert with auto break functionality.

In prototype of TMSCA, sign board is implemented using Arduino and RF transmitter as shown in the figure 1, which transmits the specific beep sound to specific type of sign board like speed breaker, school zone board etc. The vehicle also contains RF receiver and Arduino, which starts receiving the beep sound when near to sign board.Here the Arduino microcontroller is attached to the RF Transmitter which transmits radio frequency in a specified range such that it is detected by the nearby Vehicle approaching the signboard by means of an RF receiver attached to it.

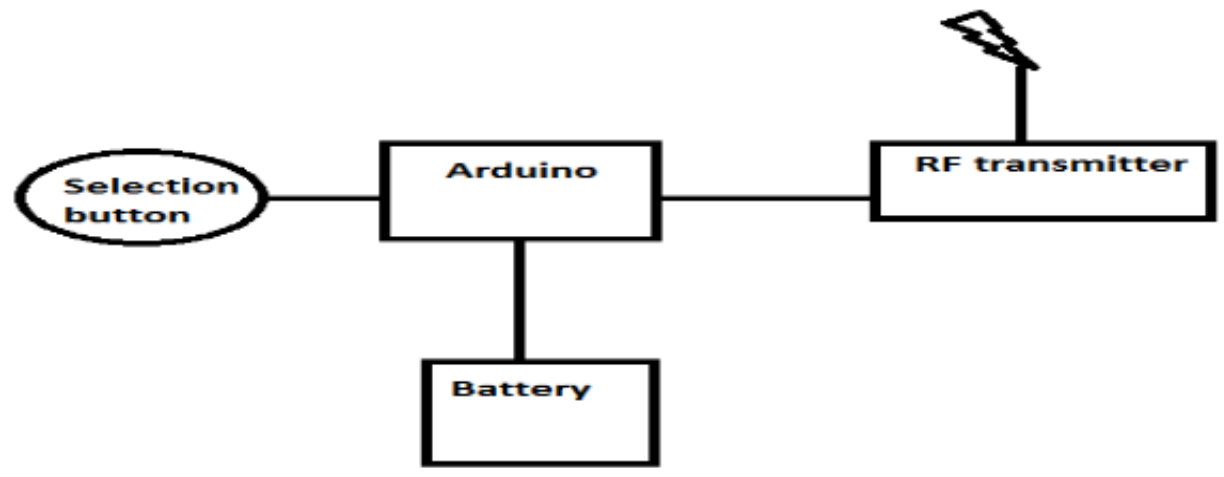

**Figure 1: Block diagram of Smart Sign Board**

After receiving the code by the approaching Vehicle as shown in figure 2, Arduino starts measuring the current speed of vehicle and if the speed is above recommended speed then it starts giving alert using buzzer. If the speed of the vehicle is not reduced even after receiving the audio alert from the buzzer, then the vehicle will auto break i.e., the vehicle speed is reduced to zero automatically. The Driver input to our prototype car is given using a Bluetooth module and mobile app.

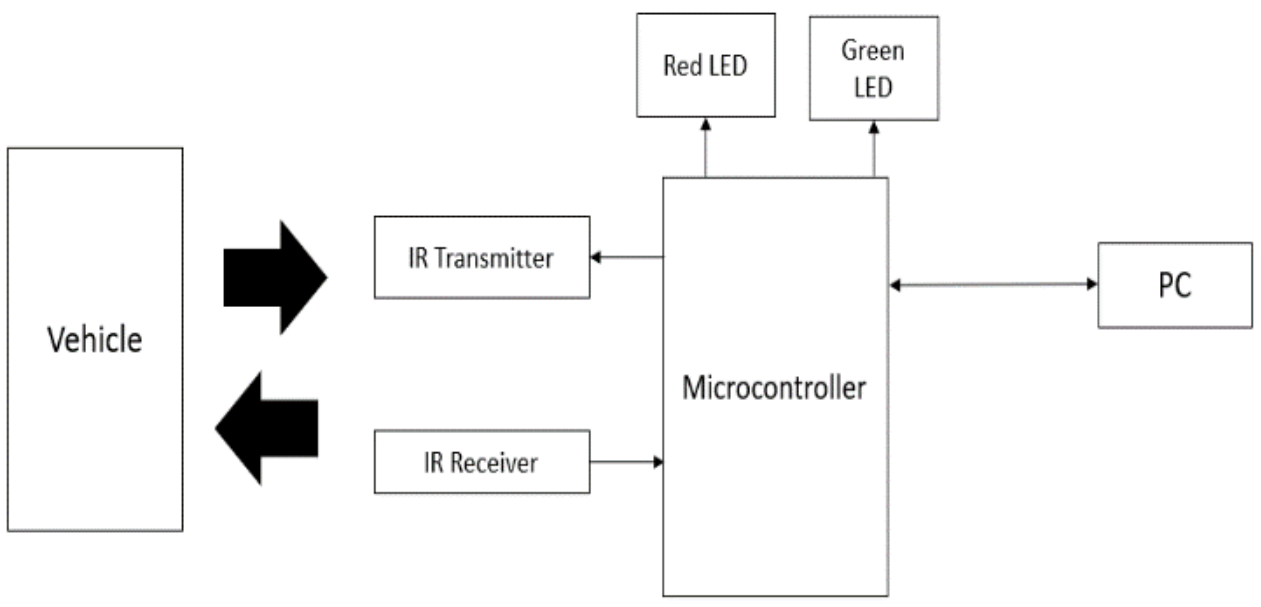

**Figure 2: Block Diagram of the Vehicle Sign Board Detection system**

As shown in Figure 3, Vehicle control system consists of Arduino microcontroller which is attached to an RF receiver. The RF receiver receives radio signal transmitted from the transmitter attached to the vehicle. The Arduino identifies the type of sign board based on the unique radio frequency emitted and selects it from the selection button.

A Bluetooth module is used for giving Driver input to the car by means of control messages, stating the vehicular speed to be reduced. After receiving the message, Arduino sends an alert by means of a Buzzer which alerts the driver to reduce the speed.

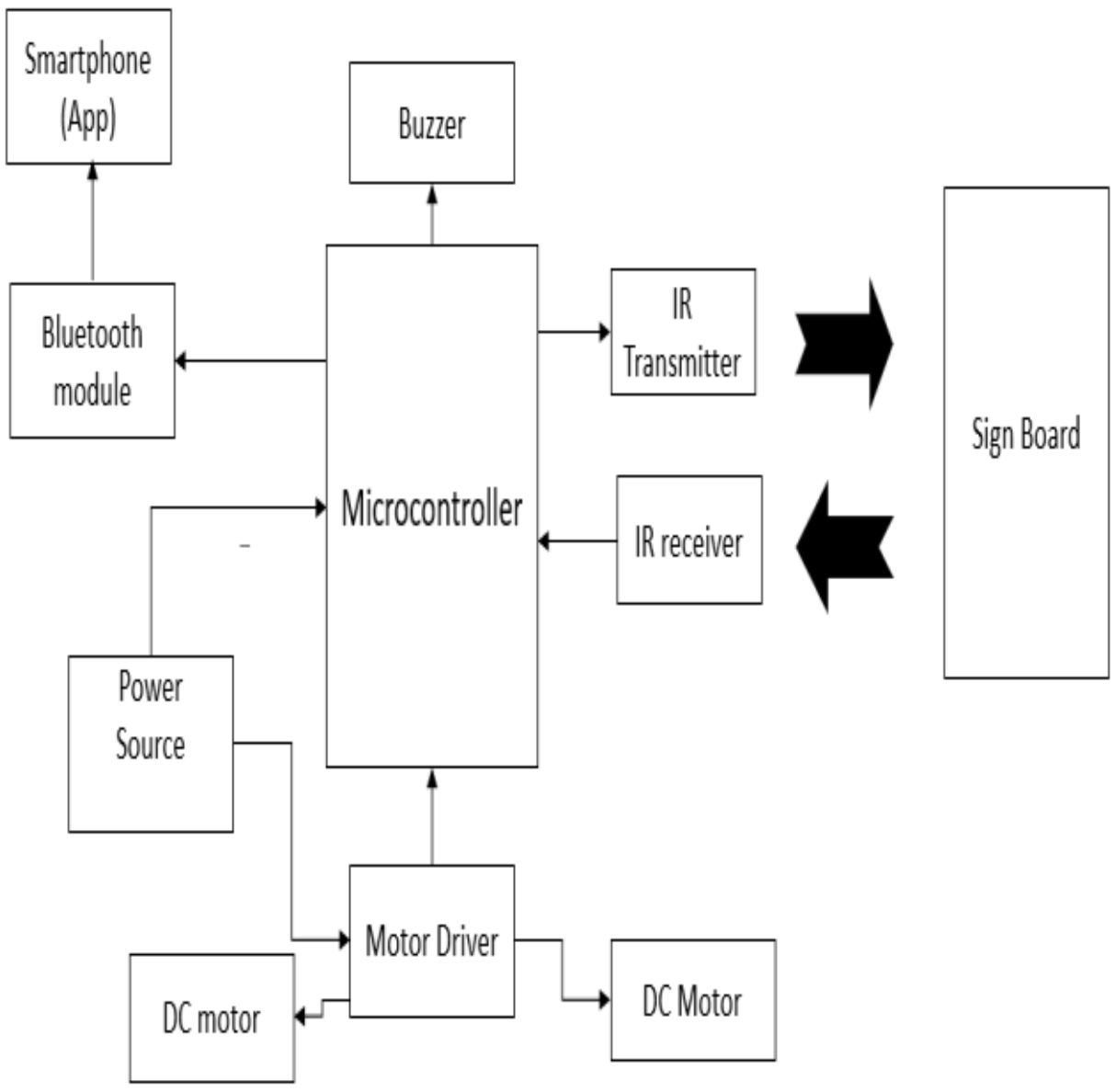

**Figure 3: Block diagram of Vehicle Control system (TMSCA)**

In case the driver fails to adjust his vehicle speed then motor of the car is brought to a halt by means of a motor-driver connected to the Arduino microcontroller. This acts as an auto-break functionality in case of emergency situations when the driver is unable to control the vehicle speed or deliberately does not follow the rules of the sign board.

**Ambulance Mode**

Whenever the emergency vehicle is detected like ambulances, Police vehicles etc.. on the lane, an algorithm is used to change the traffic light from red to green to allow the vehicle to move easily.

## 4. IMPLEMENTATION AND RESULTS

The sign board detection is composed of RF transmitter, receiver system that receives signal from the detected sign board. If the speed of the vehicle is above the recommended speed according to the sign board, then it automatically adjusts the vehicle speed to the specified speed of the sign board.

### (a) Interface between vehicle and Bluetooth system

The Bluetooth module present on the vehicle is connected to an application called Bluetooth electronics. Once the connection has been established the LED indicator will stop blinking indicating successful connection. With the help of Bluetooth module we can control the movement of vehicle ,display the mode of signboard when the vehicle reaches its vicinity or to change the signal of signboard to red or green for the emergency mode.

### (b) Communication between signboard and vehicle

The vehicle has two dc motors which are connected to the motor drivers, which is in turn connected to the microcontroller. When the microcontroller receives the signal from the signboard from the IR receiver, this input is processed and is then sent to the motor which then reduces the speed of the vehicle accordingly The signboard signal are received by the vehicle and control messages are displayed on the Bluetooth device regarding the mode of signboard and the speed to which it is set.
Thus various actuations that take place automatically in our prototype vehicle module for each particular sign board detected.

- The GPIO pins of Arduino is used to control and interface with sensors, and motors.
- Serial communication is established between Bluetooth module and vehicle (car)
- We then wait for user input (Button press). If button is pressed then car undergoes the process of movement i.e forward/backward/left/right. If the button is not pressed, the vehicle ceases to move
- We enable interrupt for serial communication between car and sign board so that when the vehicle comes within a specific range of the signboard, the microcontroller should stop all its other function and service the interrupt.
- Check for light pulses from IR transmitter. If detected, then buzzer is activated.

Then the time period of pulses is determined .Based on the time period following actions are performed.

- If pulses are between 65 -75 (humps) ,then speed is reduced to 30% speed, else
- If pulses are between 45 -55 (Speed limit) , then speed is reduced to 80% speed else
- If pulses are between 85 -95 (School zone) , then speed is reduced to 50% speed else
- If pulses are between 105 -115 (Freeway) then then speed is not restricted.

For the **Ambulance Mode**, the vehicle sends the pulses to the upcoming signboard.

- If the pulses are between 45 – 55 ms, green light of signboard is turned on i.e detected vehicle is ambulance.
- If pulses are between 65 – 75 ms red light is turned on i.e the vehicle has crossed the road.

**Algorithm**

Step 1: Start

Step 2: Initialize GPIO pins

Step 3: Select Mode for signboard.

Step 4: Check for incoming vehicle pulse signals.

Step5: If speed of vehicle greater than the set speed of signboard, alert user by buzzer and send pulse signal to restrict speed.

Step 6: Else if vehicle is emergency mode, change the signal of signboard according to vehicle input.

Step 7: Repeat in loop until user exits

Step 8: End

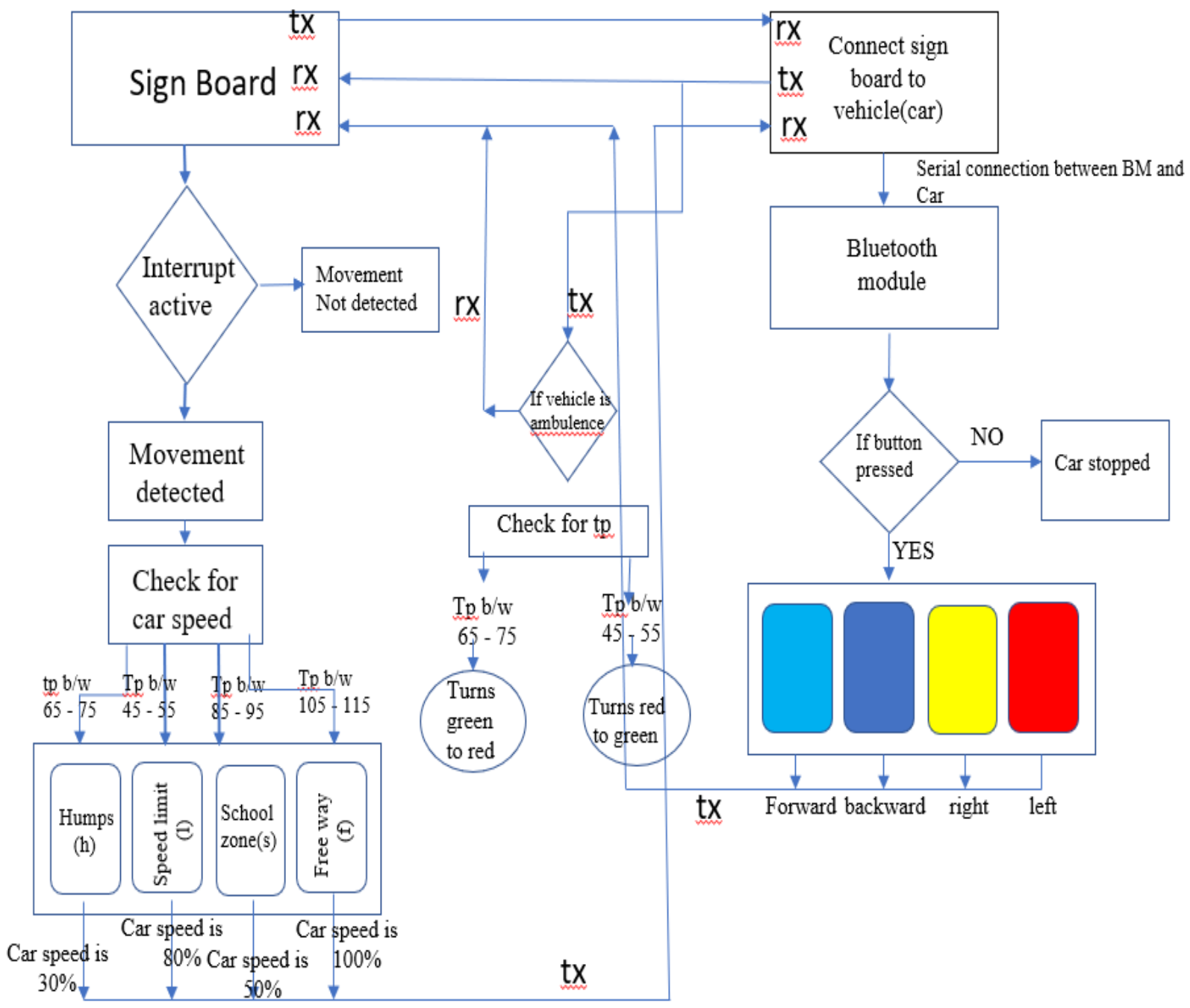

**Figure 4 : Flowchart of Vehicle speed control in TMSCA system**

Results obtained are displayed as in the figures below. It is observed that Bluetooth Module has captured four different sign boards that have been detected, namely:

i. Freeway (no sign board detected)

ii. Humps

iii .School zone and

iv. Speed Limit sign board.

The four buttons: Red, Green, Blue and Yellow are used to control the movement of the vehicle i.e., for Right Turn, Forward, Left Turn and Backward respectively.

By pressing the A and X buttons the system can be configured for ambulance mode to give red to green signals for ambulance movement on a particular lane.

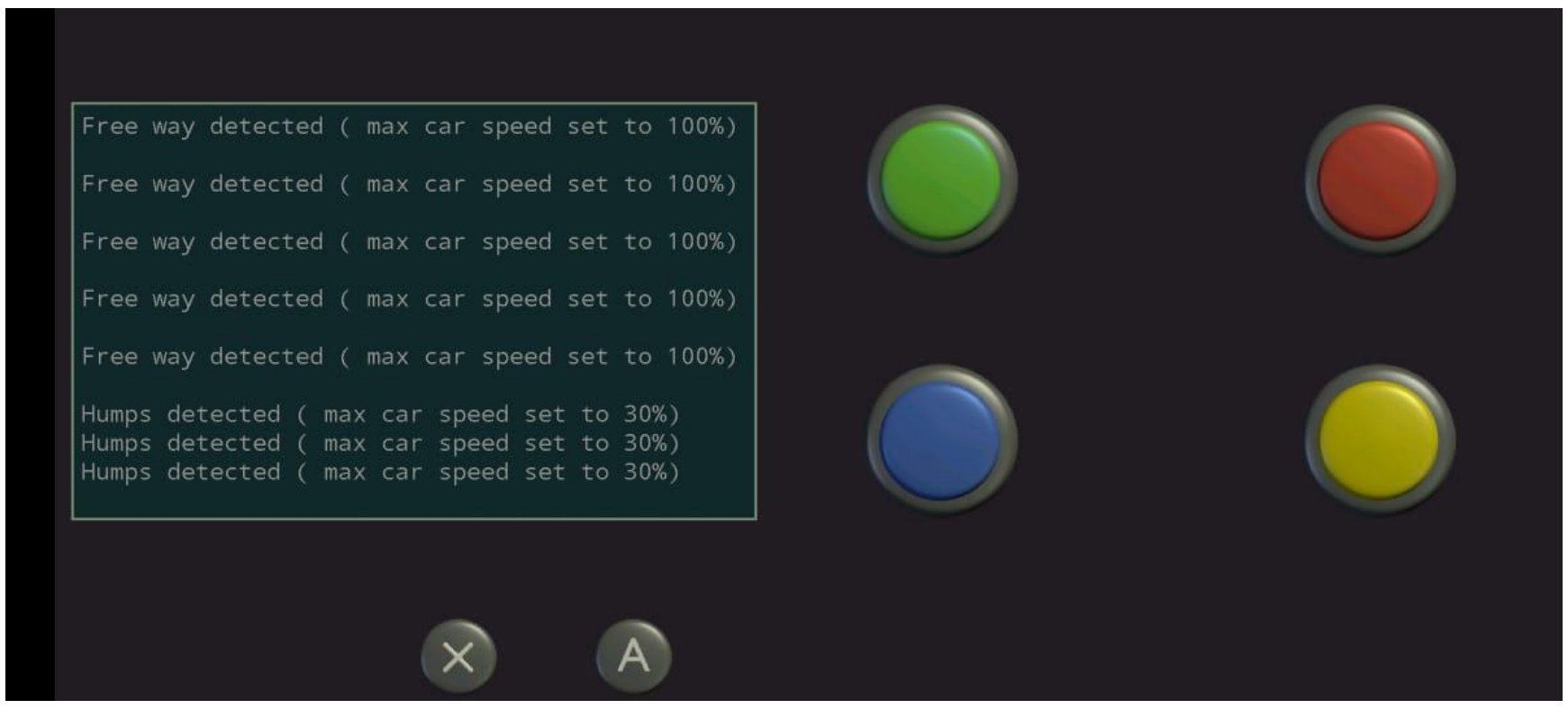

**Figure 5: Detection of Freeway and humps sign boards**

First instance of Figure 5 represents the Freeway (no sign board) detected and the car speed is automatically set to maximum. Similarly for Humps detected, the speed of car is set to 30% of maximum speed.

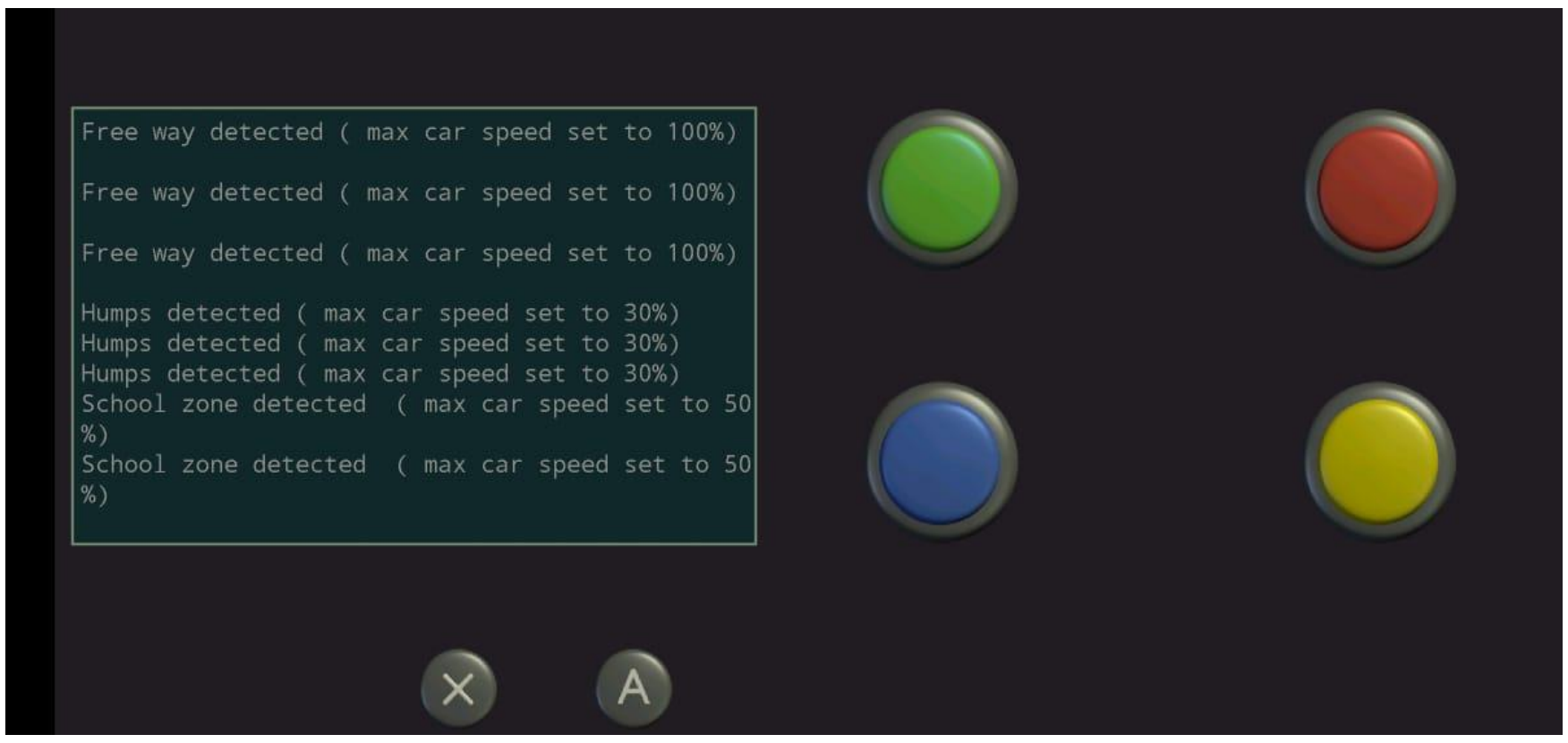

**Figure 6: Detection of School Zone sign board**

In figure 6, Sign board being detected is School Zone and accordingly the speed of car is reduced to 50% of maximum speed.

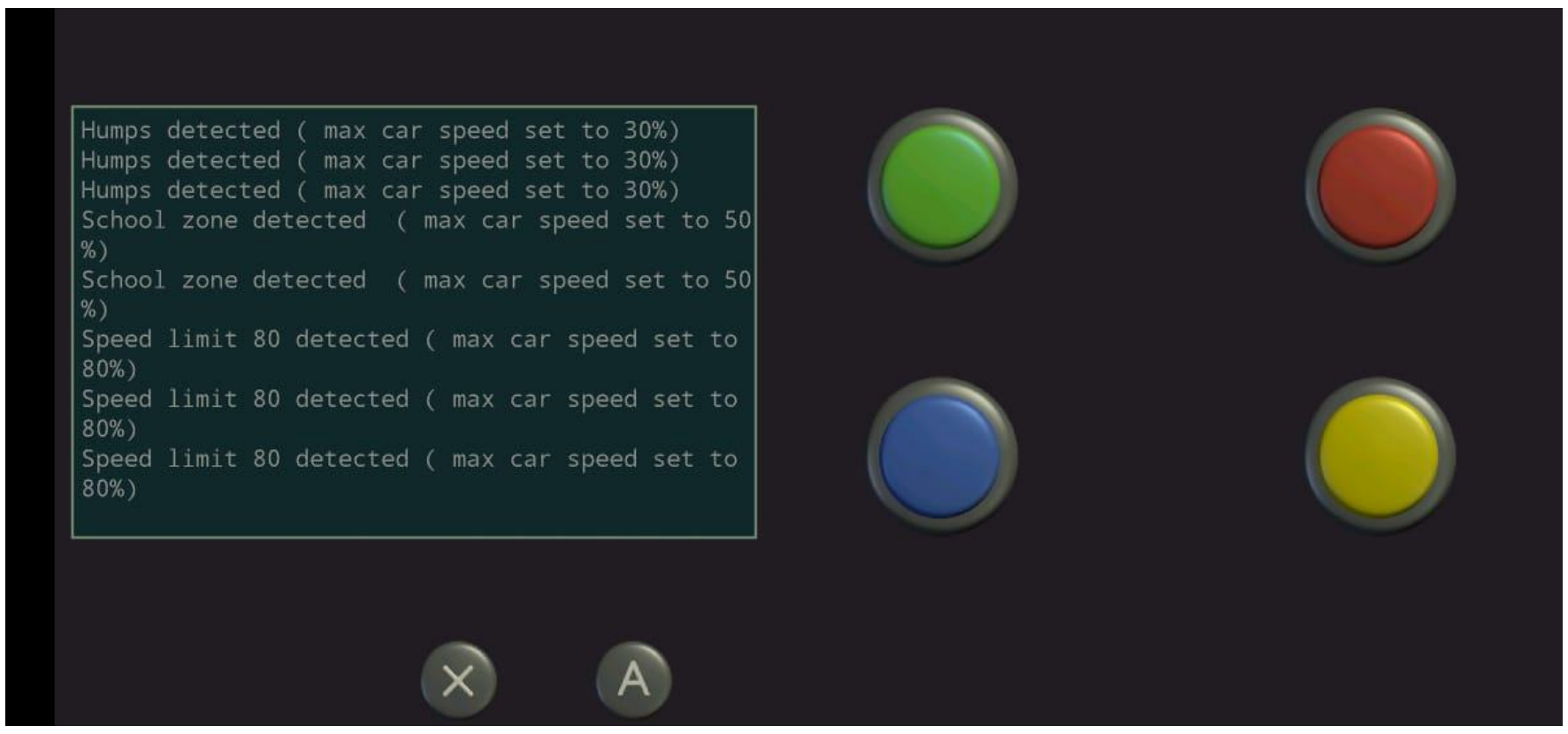

**Figure 7: Detection of Speed Limit sign board**

In figure 7, next sign board detected is the Speed Limit sign. After detecting this Sign Limit board, the speed of the vehicle is automatically reduced to 50% of its maximum speed.
All the actuations that take place in TMSCA system can be summarized as given by Table 1 shown below:

**Table 1: Sign Board detection and Speed reduction of vehicle**

| Time period of signal pulses received (Tp in mS) | Type of Sign board detected | Automatic change in vehicular speed (in terms of % of max. speed) |
|---|---|---|
| 45<Tp<55 | Speed Limit | 80% of Max. |
| 65<Tp<75 | Humps | 30% of Max. |
| 85<Tp<95 | School Zone | 50% of Max. |
| 105<Tp<115 | Freeway | 100% of Max. |

## 5. CONCLUSION AND FUTURE SCOPE

The Traffic Management and Collision Avoidance system has been carefully designed in accordance with the user requirements such as:

- Detection of sign board
- Alert the driver if speed not reduced and also auto break
- Should be able to detect various sign boards and process the required speed limit of vehicles.
- The system should be able to give preference to movement of ambulances.

In conclusion, TMSCA was specifically designed in order to combat the road accidents occurring in India on a daily basis. This objective was achieved by developing a system that would monitor the driver's vehicle and send alerts if they did not abide by the rules of any particular sign boards on the road.

This project design is also useful for drivers operating emergency vehicles like ambulances which require by passing of red traffic signals in case of emergency situations.

TMSCA is a fully automatic prototype system and will save distracted driver from getting into accident. It increases pedestrian safety. It can save many thousands of lives. It is cost effective, hence can be implemented economically.

As future enhancement the following updation can be incorporated to TMSCA system. Designing an IoT cloud database that records the drivers who do not respond to alert messages conveyed by the sign board detection system.

Record the number of special vehicles like ambulances that require traffic signal light change from red to green in case of emergencies in a congested traffic lane.